\documentclass[twocolumn]{aastex631}

\graphicspath{{./}{figures/}}
\graphicspath{{./}{figs/}}

\usepackage{units}
\usepackage[noabbrev]{cleveref}

\newcommand{\msun}{$M_\odot$}

\defcitealias{Duffell+MacFadyen+2015}{DM15}
\newcommand\DM{\citetalias{Duffell+MacFadyen+2015}}
\begin{document}

\title{Stars Bisected by Relativistic Blades}

\author[0000-0003-3356-880X]{Marcus DuPont}
\author[0000-0002-0106-9013]{Andrew MacFadyen}
\affiliation{Center for Cosmology and Particle Physics, New York University \\
New York, NY 10003, USA}



\begin{abstract}
We consider the dynamics of an equatorial explosion powered by a millisecond magnetar formed from the core collapse of a massive star. We study whether these outflows --- generated by a priori magneto-centrifugally-driven, relativistic magnetar winds --- might be powerful enough to produce an ultra-relativistic blade (``lamina'') that successfully carves its way through the dense stellar interior. We present high-resolution numerical special-relativistic hydrodynamic simulations of axisymmetric centrifugally-driven explosions inside a star and follow the blast wave propagation just after breakout. We estimate the engine requirements to produce ultra-relativistic lamina jets and comment on the physicality of the parameters considered. We find that sufficiently collimated --- half-opening angle $\theta_r \leq 0.2^\circ$ --- laminas successfully break out of a compact progenitor at ultra-relativistic velocities ($\Gamma_{\rm core} \gtrsim 30$) and extreme isotropic energies ($E_{k,\rm iso} \sim \unit[5 \times 10^{52}]{erg}$) within a few percent of the typical spin-down period for a  millisecond magnetar. The various phases of these ultra-thin outflows such as collimation shocks, Kelvin-Helmholtz instabilities, and lifetime are discussed and we speculate on the observational signatures echoed by this outflow geometry. 
\end{abstract}



\section{Introduction}\label{sec:intro}
Highly magnetized neutron stars as sources of classical gamma-ray bursts (GRBs) has been a topic spanning many decades \citep[e.g.,][]{Usov+1992,Cheng+1993,Thompson+1994,Komissarov+2004,Thompson+Chang+Quat+2004,Thompson+2005,Bucciantini+2012,Metzger+2015,Bugli+2020}. It is thought that with large enough surface magnetic fields ($B_{\rm} \geq \unit[10^{15}]{G}$) and short spin-down times, magnetars can theoretically meet the extreme energy requirements necessary to power GRBs. This is in slight contrast with another proposed GRB engine, wherein a black hole (BH) accretes several solar masses worth of stellar material during a catastrophic collapse before eventually powering a fireball that blasts its way through the stellar envelope \citep[e.g.,][]{Paczynski+1986, Paczynski+1998, Goodman+1986,Eichler+1989,Mochkovitch+1993, Woosley+1993,MacFadyen+1999}. 

While the exact dynamics preceding core collapse remains poorly understood, there is common ground on the asymptotic outflow geometry. The most common assumption promulgated about these types of explosions is that the fire ball is collimated into a relativistic, conical jet (``classical'' jet hereafter) that burrows through the collapsing star along the symmetry axis of the compact central engine. These classical jets are well supported by GRB afterglow observations \citep[see, e.g.,][for comprehensive compilations of classical GRB observations]{Kann+2010,Kann+2011,Fong+2015}, but in most cases these data are only indicative of the asymmetry of the relativistic outflow, and the exact geometry is not entirely constrained \citep[e.g.,][]{Granot+2005,Dupont+2023}. Therefore, we may rightly ask: can an equatorial jet slice its way through the dense core of a dying star like its classical counterpart? To investigate, we invoke an a priori axisymmetric millisecond magnetar (MSM) central engine that outputs a highly collimated outflow near the stellar equator.  This is admissible since it has been shown that Poynting flux-dominated flows --- as is the case for MSMs --- can be efficiently collimated \citep{Vlahakis+2003, Komissarov+2007,Bucciantini+2011},  and highly relativistic, energetic, equatorial winds can exist for pulsars like Crab \citep{Komissarov+2004,Spitkovsky+2004}. We dub these types of jets ``lamina'' jets (or ``blades'' colloquially) because of their ultra-thin resultant outflow. Better understanding of these blast wave geometries lends itself to more stringent interpretations of transients seen by ongoing and upcoming surveys \citep{Barthelmy+2005,Shappee+2014,Chambers+2016,Ivezi+2019,Bellm+2019}. In this Letter, we present a 2D axisymmetric special relativistic simulation of a lamina jet slicing its way through an 18 $M_\odot$ pre-supernova helium star and track the jet's evolution until just after breakout.

This Letter is organized as follows: Section \ref{sec:setup} discusses the numerical setup and initial conditions, in Section \ref{sec:results} we present our results, Section \ref{sec:discussion} discusses the relevance of our work, and Section \ref{sec:summary} provides a summary.

\section{Numerical Setup}\label{sec:setup}
\subsection{Governing Equations}
The governing equations in this setup are the standard special-relativistic hydrodynamic equations:
\begin{eqnarray}
(\rho u^\mu)_{,\mu} &=& \Psi\label{eq: jmu}\\
(T^{\mu\nu})_{,\nu} &=& \Theta^\mu\label{eq: tmunu_source},
\end{eqnarray}
%
where $\rho$ is proper fluid density, $[u^\mu] = \Gamma (1, \vec{\beta})$ is four-velocity, $\Gamma = (1 - \beta^2)^{-1/2}$ is Lorentz factor, $\beta$ is velocity in units of the speed of light, $c$, which is unity in our setup, $\Theta$ and $\Psi$ are source terms, and $T^{\mu \nu}$ is the stress-energy tensor for a perfect fluid,
\begin{equation}\label{eq: stress_energy}
    T^{\mu \nu} = \rho h u^\mu u^\nu + p \eta^{\mu \nu},
\end{equation}
where $h = 1 + \varepsilon + p / \rho$ is total specific enthalpy, $\varepsilon$ is internal energy, $p$ is pressure, and $\eta^{\mu \nu}$ is the Minkowski metric with signature $(-, +, +, +)$. The
set of Equations \ref{eq: jmu} -- \ref{eq: stress_energy} become closed when choosing an ideal gas equation of state $p(\varepsilon) = (\hat{\gamma} - 1)\rho\varepsilon$ where $\hat{\gamma} = 4/3$ is the ratio of specific heats at constant pressure and volume.
\subsection{Engine Model}
The source terms are modeled as Dirac delta distributions such that the power density of the engine has the form
\begin{equation}\label{eq: power_density}
    \Theta^0(\vec{r},t) = \frac{L_{\rm eng}}{2\pi r^2} \delta(r - r_n)\delta(\mu - \mu_n)g(t),
\end{equation}
where $L_{\rm eng}$ is the engine power integrated over the entire sphere, the Dirac deltas are written as Gaussian approximations:
\begin{equation}\label{eq: radial_dirac}
    \delta(r - r_n) \approx \frac{r}{r_n^2}e^{-r^2/2r_n^2},
\end{equation}
\begin{equation}\label{eq: angle_dirac}
    \delta(\mu - \mu_n) \approx \frac{f}{\sqrt{\pi}}e^{-f^2(\mu - \mu_n)^2} ,
\end{equation}
where $r_n$ is the effective radius of the engine nozzle, $\mu = \cos\theta$, $\mu_n$ is the direction of the beam, and $f$ is the geometric factor of the jet --- i.e., $f = \theta_0^{-1}$ for a lamina jet while it is $\theta_0^{-2}$ for a classical jet where $\theta_0$ is the injection angle. The function $g(t)$ is taken to be a sigmoid decay,
\begin{equation}\label{eq: decay_func}
    g(t) = \frac{1}{1 + e^{\xi(t - \tau)}},
\end{equation}
where $\tau$ is the engine duration and $\xi$ is the sharpness of the drop-off. The remaining source terms are constructed from Equation \ref{eq: power_density}, where the radial momentum density source term is
\begin{equation}\label{eq: mom_source}
    \Theta^1 = \Theta^0\beta_0 = \Theta^0 \sqrt{1 - 1/\Gamma_0^2},
\end{equation}
and the baryon loading term is
\begin{equation}\label{eq: dens_source}
    \Psi = \Theta^0 / \eta_0,
\end{equation}
where we define $\Gamma_0$ as the injected Lorentz factor and $\eta_0 \equiv \dot{E} / \dot{M}_0$ as the radiation to baryon ratio where $\dot{E}$ is the energy outflow rate near the engine and $\dot{M}_0$ is the initial mass outflow rate\!
\footnote{Some texts call $\eta$ the dimensionless entropy or the initial random Lorentz factor.}.
The engine duration can be set by requiring $\tau_{\rm bo} < \tau_{*}$ where $\tau_*$ is the spin-down time of the magnetar,
\begin{eqnarray}
    \tau_* &=& - \frac{\omega}{\dot{\omega}} = \frac{2E_{\rm rot}}{L_*} 
    \\
    &\sim& \unit[200]{s} \left(\frac{M_{\rm PNS}}{1.4 M_\odot}\right)\left(\frac{R_{\rm PNS}}{\unit[12]{km}} \right)^{-4} \left(\frac{B}{\unit[10^{15}]{G}}\right)^{-2} \left( \frac{T}{\unit[1]{ms}}\right)^2\nonumber,
\end{eqnarray}
 $\tau_{\rm bo}$ is the breakout time, $\omega$ is the rotational frequency, $E_{\rm rot}$ is the rotational energy, $L_*$ is the spin-down luminosity, $M_{\rm PNS}$ is the proto-neutron star mass, $R_{\rm PNS}$ is the proto-neutron star radius, $B$ is the surface equatorial magnetic field, and $T$ is the rotation period. In reality, it is not as simple as setting $\tau < \tau_*$ because the engine must do considerable work to displace enough stellar material to launch a successful jet. The breakout time of the beam is $\tau_{\rm bo} = \Gamma_{\rm ej} R / u_{\rm ej}(r)$, where $R$ is the stellar radius.  We can compute $u_{\rm ej}(r)$ by noting the isotropic luminosity of the jet as in  \citep{Meszaro+Waxman+2001},
\begin{equation}
    L_{\rm iso} = 4\pi r^2 u_{\rm j}^2h_j \rho_{\rm j},
\end{equation}
and balancing the pressure of the jet head with that of sub-relativistic ejecta ahead of it, i.e., $u_{\rm j}^2h_j \rho_{\rm j} = u_{\rm ej}^2 \rho_{\rm ej}(r)$, giving 
\begin{equation}
      u_{\rm ej}^2 = \frac{L_{\rm iso}}{4\pi r^2\rho_{\rm ej}(r)} = \frac{q(\theta) r L_{*}}{3M_{\rm ej}},
\end{equation}
where $q \equiv 4\pi / \Omega$, $\Omega$ is the solid angle, and  we've made use of the fact that $L_{\rm iso} = q L_{*}$. The breakout time is therefore, $\tau_{\rm bo} \sim 3\Gamma_{\rm ej}u_{\rm ej}R M_{\rm ej} / qrL_*$. Now, we revisit $\tau_{\rm bo} < \tau_*$:
\begin{equation}\label{eq: time_constraint}
    \frac{3\Gamma_{\rm ej} u_{\rm ej} M_{\rm ej} R}{q r L_*} < \frac{2E_{\rm rot}}{L_*} = \frac{2u_{\rm ej}^2 M_{\rm ej}}{L_*},
\end{equation}
where the last equality stems from assuming the rotational energy is extracted with perfect efficiency. Equation \ref{eq: time_constraint} requires $q^{-1} < \frac{2r}{3R}\beta_{\rm ej}$ for a successful breakout. \cite{Quataert+2012} compute the half-opening angle constraint for a classical jet (i.e., $q = 2 / \theta_j^2$) to be $\theta_j < \theta_{j, \rm max} \equiv (\beta_{\rm ej} / 2 )^{1/2}$. This fixes $r/R = 3/8$ in our framework. The constraint on the half-opening angle for the lamina is then
\begin{equation}\label{eq: theta_constraint}
    \theta_r < \theta_{r, \rm max} \equiv \beta_{\rm ej} / 4 =  \theta_{j, \rm max}^2 / 2,
\end{equation}
which implies that the lamina must be much more collimated than a classical jet with the same power in order to break out of the star. 
\subsection{Initial Conditions}
To set the engine power, we assume a ``split monopole'' \citep{Weber+Davis+1967} field geometry and scale the MSM luminosity calculated by \citet{Thompson+Chang+Quat+2004}, $L_{\rm eng} \approx L_{*}\vert_{t=0} = \unit[2.4 \times 10^{51}]{erg~ s^{-1}} (R_{\rm PNS}/\unit[12]{km})^{8/3}$. We also assume a relativistic, hot fireball component with $\eta_0 = 1000$, consistent with a radiatively driven engine\!
\footnote{In reality $\eta$ and other parameters are time variable, but since $L_{\rm *} \propto (1 + t/\tau_*)^{-2}$ we can set the engine duration to $\tau = \epsilon \tau_*$ where $\epsilon \ll 1$. This would ensure $|\Delta L_* / L_*| \sim 2\epsilon$ is small enough to assume rough constancy in our simulations.}.
In a scenario in which the magnetar is accreting, the spin-down is counteracted by angular momentum transport from the fallback material \citep[e.g.,][]{Parfrey+2016,Metzger+2018}, so we assume rough constancy within the timescale considered in this work by setting $\xi = 100$ in Equation \ref{eq: decay_func}. We do not explicitly invoke magnetic fields since we assume all of the magnetic energy outside of the magnetar light cylinder is converted into kinetic energy, e.g., by magnetic dissipation, \citep[see][]{Spruit+2001, Vlahakis+2003, Komissarov+2009}. Our engine-progenitor model is reminiscent of \citet[hereafter \DM]{Duffell+MacFadyen+2015} wherein they launch a successful collapsar jet through a star with $\theta_{0, \rm DM15} = 0.1$. With this classical jet injection angle as a baseline, we use Equation \ref{eq: theta_constraint} to set $\theta_0 = 0.005$ to match the power density of the classical jet. For an equatorial outflow, we set $\mu_n = 0$ in Equation \ref{eq: angle_dirac}. The stellar model is an 18 {\msun} Wolf-Rayet star that was originally a 30 {\msun} Zero Age Main Sequence (ZAMS) star rotating at 99\% breakup. The progenitor was evolved using the Modules for Experiments in Stelar Astrophysics \citep[MESA;][]{Paxton+2011,Paxton+2013,Paxton+2015,Paxton+2018,Paxton+2019} code, and we invoke the density profile for this star fitted by \DM,
\begin{equation}\label{eq: density_profile}
    \rho(r) = \frac{\rho_c \times \max \left(1 - r / R_3, 0\right)^n}{1 + (r / R_1)^{k_1} / \left[1 + (r / R_2)^{k_2}\right]} + \frac{A}{r^2},
\end{equation}
where $A = A_* \dot{M}/4\pi v_{\rm wind} = A_* \times \unit[5 \times 10^{11}]{g ~cm^{-1}}$ is the ambient medium mass-loading parameter and the remaining parameters are defined in Table \ref{tab:params}. We use an axisymmetric spherical-polar grid with logarithmic radial zones and uniform angular zones. We enforce 1024 radial zones per decade and $\delta \theta = \theta_0 / N_{\rm beam}$, where $N_{\rm beam}$ is the number of zones within the half-opening angle of the beam. We fix $N_{\rm beam} = 10$ in our simulations. The domain range is $r\in[0.001,10] R_\odot$ and $\theta \in [0, \pi/2]$, which corresponds to 4096 radial zones by 3142 angular zones. The initial pressure and velocity everywhere are negligible. All variables are made dimensionless through combinations of fundamental constants: $R_\odot$, $c$, and $M_\odot$. This concludes the initial conditions required to launch the relativistic lamina jet into the stellar progenitor. The problem is simulated using an open source GPU-accelerated second-order Godunov code entitled \texttt{SIMBI} \citep{simbi}, written by this Letter's first author. It uses a piecewise linear reconstruction algorithm to achieve second-order accuracy in space and second-order Runge-Kutta is employed for the time integration. $\theta_{\rm PLM}$, a numerical diffusivity parameter for second order schemes, is fixed to 1.5 in our simulations for more aggressive treatment of contact waves. 
\begin{deluxetable}{lll}[t!]
\tablenum{1}
\tablecaption{Stellar Model \& Engine Parameters\label{tab:params}}
\tablewidth{0pt}
\tablehead{
Variable & Definition & Value\\
}
\startdata
$\rho_\odot$ & Characteristic density scale & $M_\odot / R_\odot^3$ \\
$t_\odot$ & Characteristic time scale & $R_\odot / c$ \\
$L_0$ & Characteristic power scale & $M_\odot c^2 / t_\odot$ \\
$\rho_c$ & Central density & $3 \times 10^7 \rho_\odot$  \\
$\rho_{\rm wind}$ & Wind density at surface & $10^{-9} \rho_\odot$ \\
$R_1$ & First break radius & 0.0017 $R_\odot$ \\
$R_2$ & Second break radius & 0.0125 $R_\odot$ \\
$R_3$  & Outer radius & 0.65 $R_\odot$\\
$k_1$  & First break slope & 3.25\\
$k_2$  & Second break slope & 2.57 \\
$n$      & Atmospheric cutoff slope & 16.7 \\
\hline
\hline
$\Gamma_0$ & Injected Lorentz factor & 10 \\
$\eta_0$ & Initial radiation-to-baryon ratio & 1000 \\
$L_{\rm eng}$  & Engine power & $3.2 \times 10 ^{-3} L_0$\\
$\tau$  & Engine duration & $2~t_\odot$\\
$\theta_0$  & Engine half-opening angle & 0.005 \\
$r_n$      & Nozzle radius & $0.01~R_\odot$ \\
$A_*$      & Dimensionless wind parameter & 1 \\
\enddata
\end{deluxetable}
%
\section{Results}\label{sec:results}
\subsection{A bisected Wolf-Rayet star}
\begin{figure}[t]
\centering
\vspace*{-4.5em}
\gridline{\fig{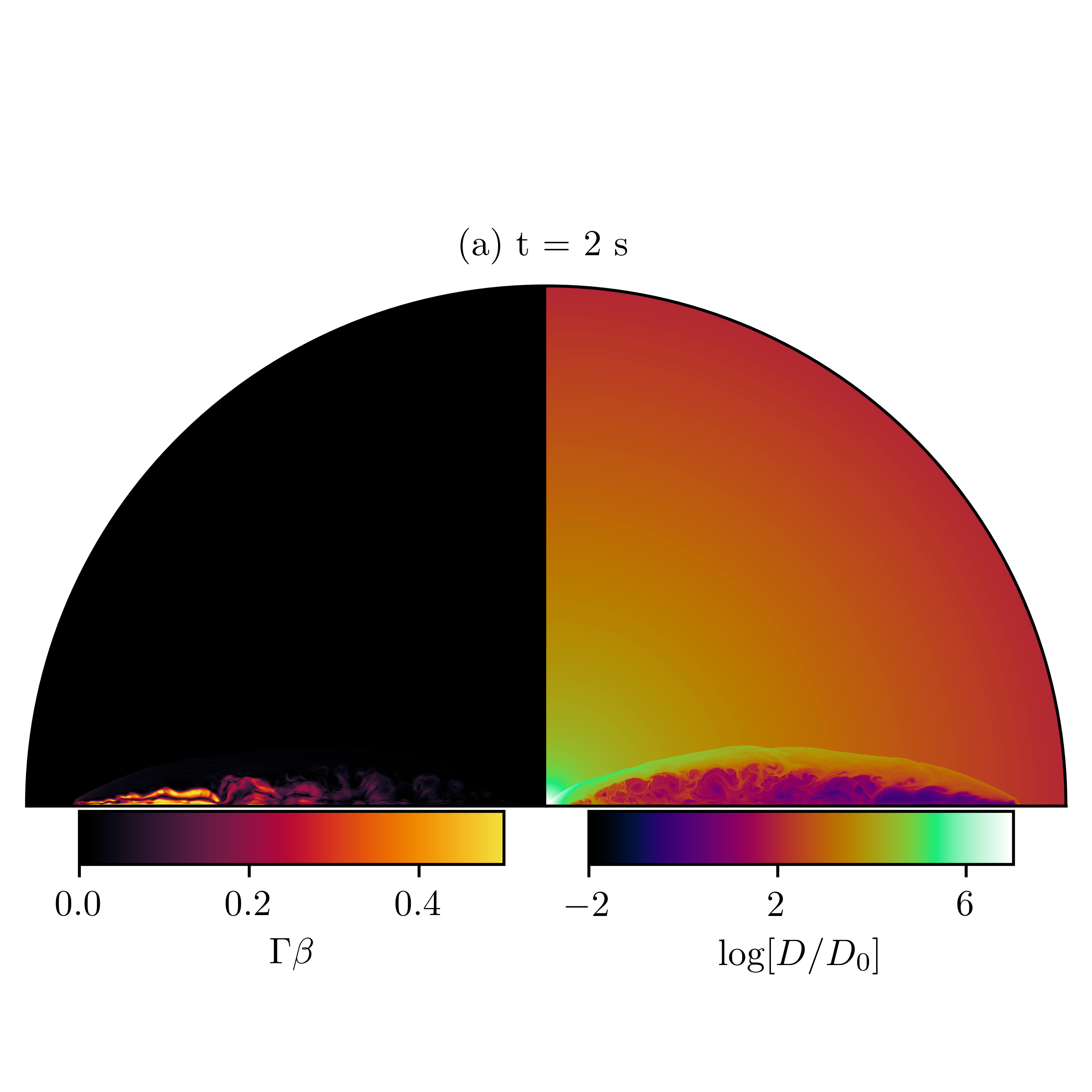}{\columnwidth}{}}\vspace*{-8.5em}
\gridline{\fig{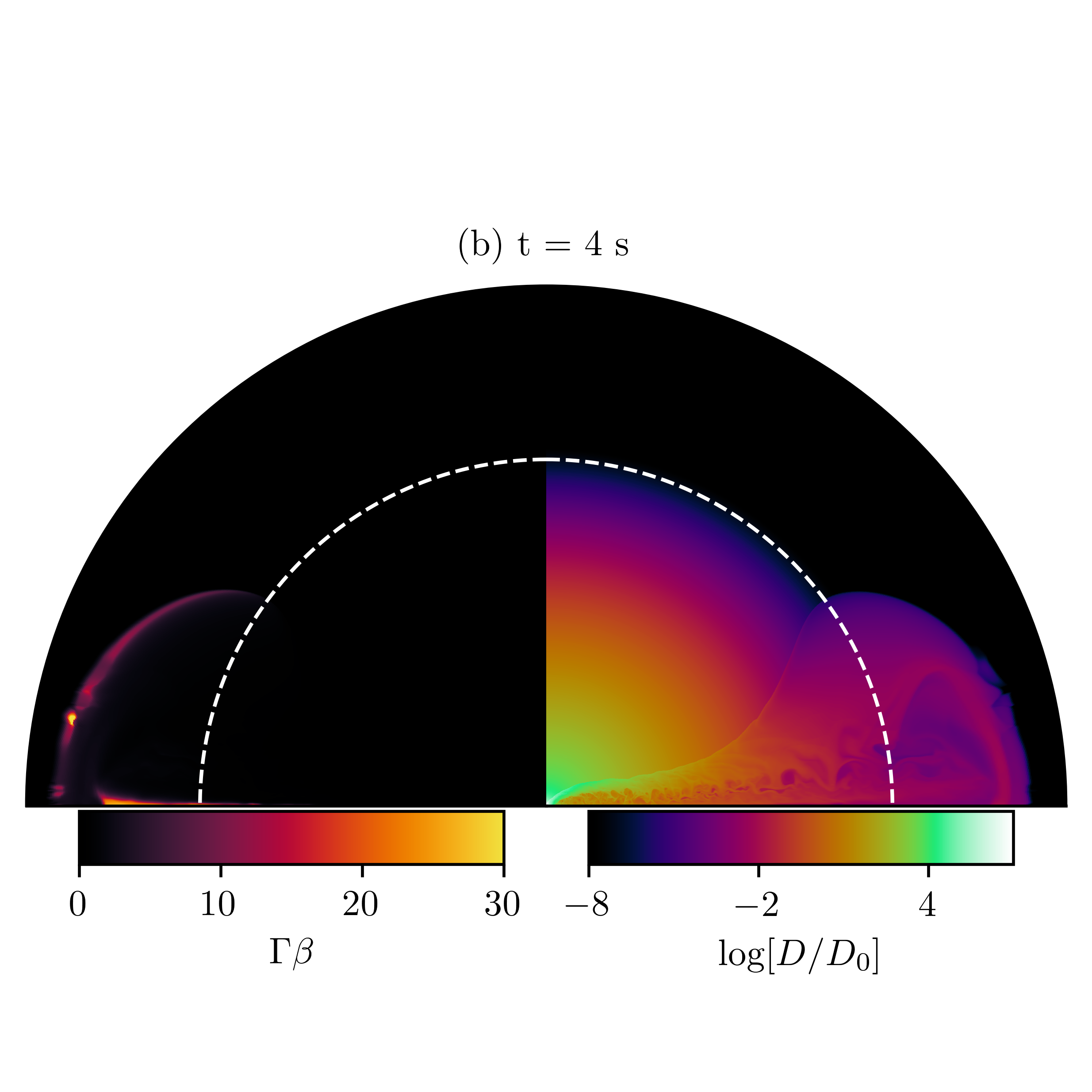}{\columnwidth}{}}\vspace*{-3.5em}
\caption{Snapshots of the lamina jet with injection angle $\theta_0 = 0.005$. The left half shows the four-velocity and the right shows the lab frame density $D = \Gamma \rho$. Panel (a) shows the interior beam zoomed in to $0.1R_\odot$ at the boundary. We see the development of Kelvin-Helmholtz instabilities as the beam tunnels through the dense core with the region nearest the jet head experiencing a few collimation shocks.  In panel (b), the lamina breaks out of the star successfully with a $\Gamma \geq 30$. The white-dashed line marks the radius of the progenitor at $0.5R_\odot$.}
\label{fig: breakout}
\end{figure}

Our fiducial runs show promising relativistic breakout for highly collimated lamina jets. Figure \ref{fig: breakout} shows the time-evolved snapshots of the explosion from the initially stationary conditions to the relativistic breakout of the beam. Since the lamina is ultra-thin and radiative, it can easily push aside matter as the beam tunnels through the dense stellar interior, allowing it to get out within four light crossing times of the progenitor.\@ Kelvin-Helmholtz instabilities develop deep in the interior as the beam propagates through the very thin cocoon, and the the jet core is naked once it accelerates down the steep density gradient ahead of it. The maximum Lorentz factor increases monotonically and exceeds the injected value, $\Gamma_0$. This hints at the fact that GRBs might not be very sensitive to the intricacies of their complicated central engines outside of a terminal bulk Lorentz factor $\Gamma_\infty \sim \eta$ \citep{Zhang+2003,Meszaros+2006}. In just a few percent of a typical MSM spin-down time, the beam breaks out at ultra-relativistic velocities before the cocoon has traversed $\sim 40$\% of the star, affecting a clean slice through the progenitor. 

%
%
\subsection{Collimation shocks}
\begin{figure}[t]
    \centering
    \centering
    \vspace*{-4.5em}
    \gridline{\fig{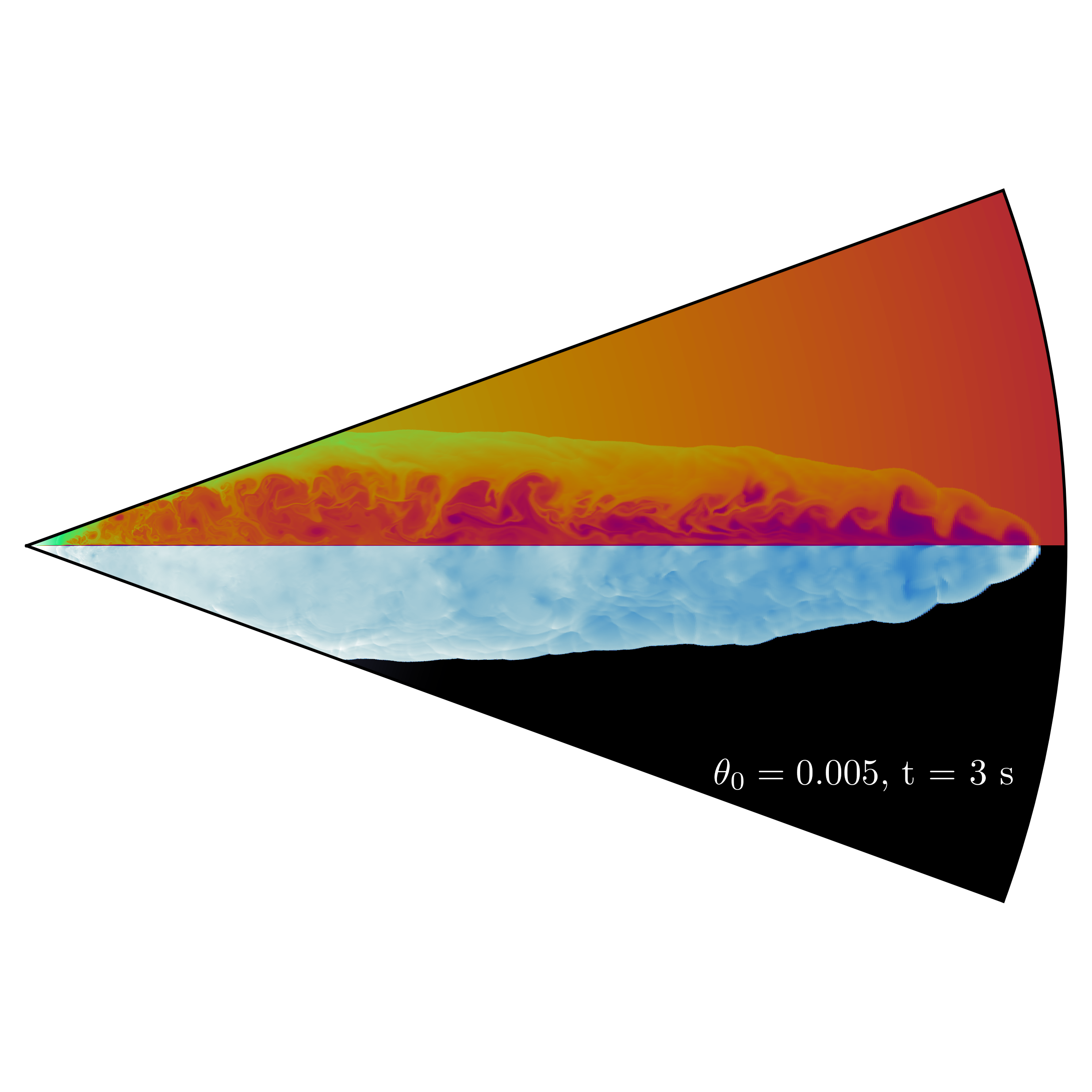}{\columnwidth}{}}\vspace*{-10.0em}
    \gridline{\fig{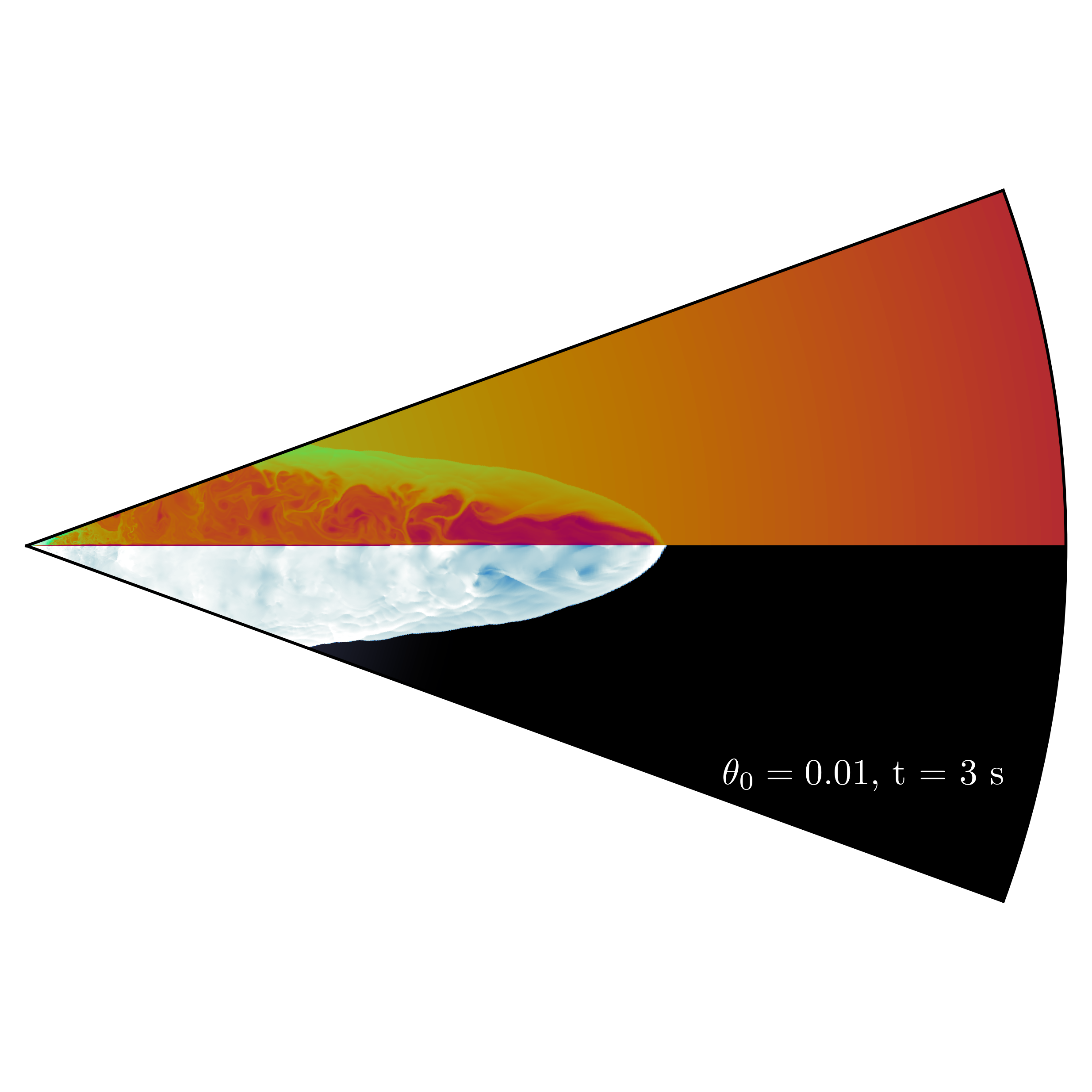}{\columnwidth}{}}\vspace*{-2.5em}
    \caption{Snapshots at $\unit[3]{s}$ depicting the lab frame density in the northern hemisphere and pressure in the southern hemisphere where the radial boundary is at $0.2R_\odot$. The density color stretch is identical to panel (a) in Figure \ref{fig: breakout}. The pressure colormap is chosen such that direction of increasing pressure gets brighter. \emph{Upper}: The beam propagation for $\theta_0=0.005$. In the equatorial plane, the beam path is evacuated and the pressure is slightly laterally stratified. \emph{Lower}: The same variables as in the upper wedge, but with injection angle $\theta = 0.01$. The cocoon tail is fatter in this instance and the jet head lags behind the thinner beam. In both cases, there exist no clear knots or pinching as usually seen for collimation shocks in classical jets.
    }
    \label{fig:density_and_pressure}
\end{figure}

Throughout its evolution, the lamina jet appears to experience collimation shocks
as evidenced by Figure \ref{fig:density_and_pressure}. In Figure \ref{fig:density_and_pressure}, we also include another lamina jet with double the opening angle $\theta_{0,\rm wide} = 0.01$ to gauge key differences in the cocoon evolution, which depicts a more advanced blast wave for a thinner beam. We believe this is due to the relatively negligible thickness of the blade-like structure of the outflow. The nature of this effect is two-pronged. That is to say, the $\theta_0 = 0.005$ beam's working surface is is half as small as the $\theta_{0,\rm wide} = 0.01$ beam, which means it is impeded by half of the mass and has double the pressure. As the thinner lamina pierces through the star, it interacts with less stellar material that is mixed in the Kelvin-Helmholtz layers of the cocoon and therefore the pressurized cocoon has a lesser impact on the collimation of the flow. Because of this, the beam encounters fewer interactions from rarefaction waves and shocks as the jet-cocoon interface is propagated throughout the star. Another way of putting it is that with the effective working surface of the engine reduced, the wave more easily travels through a star analogous to how a sharper knife more easily cuts through material while keeping the applied force fixed. We suspect that this is also the case for skinny classical jets. However, the knots from the collimation for classical jets are more prominent than for lamina jets as evidenced by previous numerical studies on classical jets \citep[e.g.,][]{MacFadyen+1999,Aloy+2000,MacFadyen+2001,Zhang+2003,Tchekhovskoy+2008,Mosta+2014,Mandal+2022}. Of course, this would have to be further analyzed in 3D to encapsulate a fuller picture of beam deflection and various instabilities that might arise.
%
%

\subsection{Outflow lifetime}
\begin{figure}[t]
    \centering
    \includegraphics[width=\columnwidth]{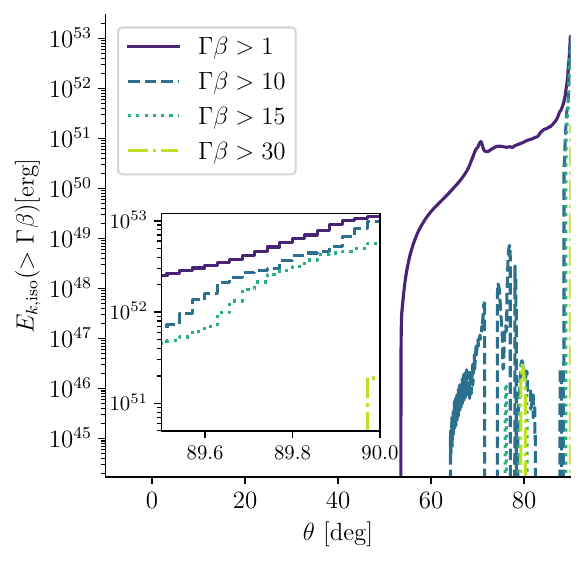}
    \caption{The cumulative isotropic-equivalent kinetic energy as a function of polar angle $\theta$ at $\unit[4]{s}$. The solid, dashed, dotted, and dash-dotted lines mark the $\Gamma\beta > \{1, 10, 15, 30\}$ cutoffs, respectively. The inset is a zoom in around some small solid angle near the stellar equatorial plane. Some of the relativistic material is forced to spread laterally as shown by the area under the $\Gamma \beta > 1$ curve, but as we discriminate towards higher velocities, the bulk lamina structure ($\Gamma \beta > 10$) is focused in the stellar equator into a thin layer with $E_{k, \rm iso} \sim \unit[5 \times 10^{52}]{erg}$. At simulation end, only a small amount of material is accelerated near the maximum Lorentz factor, with $E_{k, \rm iso}(>30) \sim \unit[2 \times 10^{51}]{erg}$.}
    \label{fig: de_domega}
\end{figure}

Although the injection angle was $\theta_0 = 0.3^\circ$, the lamina at or above $\Gamma_0 = 10$ breaks out with a half-opening angle $\theta_r \sim 0.2^\circ$
\footnote{Measured using $\sin\theta_r = E / E_{\rm iso}$}
and the carries a maximum Lorentz factor $\Gamma_{\rm core} \sim 30$ at the time of the simulation end. An observer within the stellar equatorial plane and whose line of sight passes through the lamina centroid would see $1 / \Gamma = 3\%$ of the total structure. If extreme beaming took place --- i.e., $\Gamma \gg \theta_r^{-1}$ --- the lamina would travel through the interstellar medium (ISM) with fluid parcels causally disconnected from their neighbors which would help maintain a non-spreading outflow until it sweeps up a mass $M/\Gamma$ and slows down \citep{Porth+2015}. In Figure \ref{fig: de_domega}, we compute the cumulative isotropic-equivalent energy per solid angle at simulation end,
\begin{equation}\label{eq: eomega}
    E_{k,\rm iso} (>\Gamma_{\rm c}\beta_{\rm c}; \theta) = 4\pi \frac{dE}{d\Omega}(>\Gamma_c \beta_c; \theta),
\end{equation}
where $\Gamma_{\rm c} \beta_{\rm c}$ is the four-velocity cutoff. Moreover, we estimate the motion of the bulk flow by noting the mean energy-weighted four-velocity,
\begin{equation}\label{eq: avg_four_velocity}
    \langle \Gamma \beta \rangle_{E} = \frac{\int_V \Gamma \beta (\Gamma^2 \rho h - p - \Gamma \rho)dV}{\int_V (\Gamma^2 \rho h - p - \Gamma \rho)dV},
\end{equation}
moving  above some fixed value. We are interested in the material which is moving at or above the injected Lorentz factor, $\Gamma \geq \Gamma_0 = 10$. Therefore, the ultra-relativistic component gives $\langle \Gamma \beta \rangle_E \sim 15$ as the velocity of the bulk flow. From Figure \ref{fig: de_domega} we find that $E_{k, \rm iso} (>15) = \unit[3 \times 10^{52}]{erg}$ is focused purely in the equator. This beam has mass $M(>15)|_{\theta=90^\circ} = E_{k, \rm iso} / \langle \Gamma \rangle_E \approx 10^{-3} M_\odot$. Assuming the MSM engine spontaneously shuts off, the lamina will decelerate after sweeping up $7 \times 10^{-5} M_\odot$ which, for $A_* = 1$,  occurs at a radius $r_{\rm dec} = \unit[2 \times 10^{16}]{cm}$ or $6 \times 10^5 R$. 

\subsection{Supernova energy budget}
Although the lamina breaks out of the star inefficiently (i.e., $\tau_b \sim 4R$), the cocoon has little time to traverse the remaining undisturbed star before the jet head outruns the explosion. To get a sense of the remaining energy that might be attributed to a supernova, we estimate this by summing all of the \emph{total} energy available in the slowest material: $E_{\rm T}(<0.1) \sim \unit[3 \times 10^{51}]{erg}$. Only about $30\%$ of this total energy is kinetic at the time of the simulation end. Thus, once the cocoon finishes its journey throughout the remaining stationary star and full conversion of the thermal energy into kinetic energy is complete, the energy liberated in the explosion is of order the canonical supernova explosion energy. About 10\% the deposited energy from the engine is shared with the supernova component if it were to shut off spontaneously after four seconds.

\section{Discussion}\label{sec:discussion}
A centrifugally-focused millisecond magnetar central engine at the core of a compact Wolf-Rayet star
gives rise to both the formation and the propagation of a relativistic lamina through 
the mantle and envelope of the star and to a supernova explosion. This is assuming an axisymmetric ``split-monopole'' magnetic field structure deep in the stellar interior which might fully dissipate magnetic energy due to the equatorial current sheet \citep{Spruit+2001,Drenkhahn+2002,Lyutikov+2003}. This allows for energy deposition at a rate $L_* = \unit[\text{few}\times 10^{51}]{erg~s^{-1}}$ \citep{Thompson+Chang+Quat+2004,Thompson+2005}, which is a bit larger than the rate for a typical dipole field MSM.  In just $\unit[4]{s}$ or four light crossing times of the progenitor, a clean relativistic blade-like structure bisects the helium core and breaks out into the dense circumstellar medium intact.  

\subsection{Causality and Stability}
The ring will come into causal contact with its edges when the
angle of influence, i.e., the proper Mach angle \citep{Konigl+1980}, is comparable to the angular size of the beam, viz., $\theta_{\rm M} / \theta_r \geq 1$. The relativistic Mach angle evolves as $\tan \theta_{\rm M} = u_s / u \propto 1 / \Gamma$ and angular size of the ring is $\theta_r = r_\perp / r$ where $u_s$ is the proper sound speed and $r_\perp$ is the size of the ring perpendicular to the flow. Together, these functions imply the causality constraint,
\begin{equation}\label{eq:causaility1}
    \frac{\theta_{\rm M}}{\theta_r} \propto \frac{r}{r_\perp \Gamma}.
\end{equation}
Assuming the thermal energy dominates the rest mast energy of the particles, i.e., $\rho h \propto p$, constancy of energy implies $\Gamma^2 r^2 \Delta r \theta_r p = \text{constant}$ where $\Delta r$ is the width of the annular blast wave. Energy conservation together with mass conservation, $\Gamma r^2\Delta r \theta_r \rho = \text{constant}$, leads to the scalings $\Gamma \propto \rho^{1-\hat{\gamma}}$ and $r_\perp \propto r^{-2}\rho^{-\hat{\gamma}}$ after utilizing $\Delta r \sim r / \Gamma^2$, $r\theta_r = r_\perp$, and $p \propto \rho^{\hat{\gamma}}$. Equation \ref{eq:causaility1} thus becomes 
\begin{equation}\label{eq:causaility_final}
    \frac{\theta_{\rm M}}{\theta_r} \propto r^{3 -k(2\hat{\gamma} - 1)}.
\end{equation}
Equation \ref{eq:causaility_final} implies that for an ultra-relativistic ring with $\hat{\gamma} = 4/3$, the critical ambient medium density slope is $k = 9/5$ where $k < 9/5$ implies full causality is reached while $k > 9/5$ implies causality is lost. This critical ambient medium slope is different from the $k = 2$ classical jet value calculated by \citet[]{Porth+2015}. Note that $k = 2$ is the value for stellar winds, so Equation \ref{eq:causaility_final} implies that small perturbations in the $\phi$-direction might excite interesting instability modes if the ring evolves in a wind-like environment, and we plan to address this in another paper.   Moreover, in the scenario in which the ambient medium is not uniform in $\phi$, the ring-like blast wave will become corrugated in the $\phi$ direction like pizza slices of uneven length, which might drive quasi-periodic radiation signatures in the light curves. While interior to the star and in in full 3D, the blade is likely to bend and wobble which might excite further azimuthal instabilities, which we also plan to investigate in a follow up work. 

\subsection{Visibility}
The overall blade carrying a Lorentz factor $\Gamma_{\rm beam} \geq \Gamma_0 = 10$ has half-opening angle $\theta_{\rm beam} = 0.2^\circ$, and it contains an ultra-relativistic core with $\Gamma_{\rm core} \sim 30$ and half-opening angle $\theta_{\rm core} = 0.03^\circ$ at simulation end. If the engine was instantaneously shut off from the moment of breakout, the bulk flow,  which moves with $\langle \Gamma \beta \rangle_{E} = 15$, propagates more than 5 orders of magnitude beyond the stellar radius before slowing down. To gauge when these outflows will become observable on Earth, 
we compute the photosphere radius assuming a grey atmosphere optical depth of $2/3$,
\begin{eqnarray}
    \tau = \frac{2}{3} &=& f_{\rm KN{}}\kappa_{\rm T} \int_{R_p}^\infty \rho_{\rm wind}(r) dr \\
    &\Rightarrow& R_P = A_* f_{\rm KN} \times \unit[1.5 \times 10^{11}]{cm}  \nonumber,
\end{eqnarray}
where $f_{\rm KN}$ are the Klein-Nishina corrections, $\kappa_{\rm T} =\unit[0.4 / \mu_e]{cm^2~g^{-1}}$, $\mu_e = 2/(1 + X_H) \simeq 2$ is mean molecular weight per electron, and $A = A_* \times \unit[5 \times 10^{11}]{g~cm^{-1}}$. This implies that the environment becomes optically thin within just a few stellar radii from the source, making the blade visible almost immediately post breakout.

\subsection{Observations}
In the scenario where lamina jets are sources of long gamma-ray bursts (lGRBs), their observational implications are immediately realized due to simple geometric arguments \citep{Thompson+2005,Granot+2005,Dupont+2023}. Once the jet slows down, it will spread in a single dimension, leading to a shallower break in the afterglow light curves that makes it clearly distinguishable from a classical GRB jet configuration. The characteristic afterglows in the low and high frequency bands for such rings were computed by \citet[]{Dupont+2023}{} and show that, over both frequency ranges, the ring-like blast waves are intermediate between jet-like and spherical outflows. With the afterglow decay distributions being so varied \citep[e.g.,][]{Panaitescu+2005}, we believe ring-like explosions are also candidate outflows in this arena. Note that in this work we took the extreme case of a split-monopole axisymmetric engine to achieve such relativistic outflows. A more modest dipolar engine might refocus the blade into bipolar bubbles \citep[e.g.,][]{Bucciantini+2007} if the magnetic hoop stress dominates once the magneto-centrifugal wind reaches the termination shock. Another outcome could be that if the engine power still dominates over the magnetic hoop stress, the equatorial wind might produce slower material at breakout due to having a smaller energy dissipation rate, but this likely broadens the types of transients created by the blast wave. In principle, Equations \ref{eq: jmu} -- \ref{eq: decay_func} are scale free, so rescaling opens avenues for providing explanations for transients like super-luminous supernovae, X-ray bursts, fast blue optical transients (FBOTs), and low-luminosity GRBs, to name a few. 

\section{Summary}\label{sec:summary}
We have demonstrated that an equatorial engine can produce an ultra-relativistic breakout focused into a very thin lamina structure. The fastest core is focused into an even thinner working surface due to the path ahead of the engine being evacuated so efficiently. Kelvin Helmholtz instabilities develop deep in the stellar mantle, and the cocoon-jet interface experiences rarefaction waves and/or shocks that lead to collimation shocks as usually seen for classical jets. However, the formation of ``knots''caused by the collimation shocks along the beam do not form as they have for classical jet. Because of their geometry, the lamina outflow sweeps up more mass than classical jets by a factor $\sim  \theta^{-1}$, so to achieve similar ultra-relativistic breakouts, their engines must be highly focused at the outset. In the future, we plan to address this same problem, but in three dimensions to capture any instabilities such as corrugated (accordion-like) waves that might arise in the $\phi$ direction due to causality effects or non-uniform circumstellar environments. Furthermore, evolving the lamina over thousands of decades in distance to understand the late-time geometry of the explosion once the relativistic beam slows down fully must also be done to capture distinctive features in the observational signatures of these types of outflows. We also suggest that a detailed resolution study of relativistic jet breakout is needed to further pin down the limitations imposed by these compact GRB progenitors and seeding of turbulence.

%

\begin{acknowledgments}
M.D. thanks Eli Waxman, Andrei Gruzinov, and Brian Metzger for helpful discussions, a James Author Fellowship from NYU's Center for Cosmology and Particle Physics, and the LSSTC Data Science Fellowship Program, which is funded by LSSTC, NSF Cybertraining Grant \#1829740, the Brinson Foundation, and the Moore Foundation.  We acknowledge support from NASA ATP grant 80NSSC22K0822.
\end{acknowledgments}

%





\bibliography{refs}{}
\bibliographystyle{aasjournal}



\end{document}